# In Quest of the Better Mobile Broadband Solution for South Asia Taking WiMAX and LTE into Consideration

Nafiz Imtiaz Bin Hamid, Md. R. H. Khandokar, Taskin Jamal, Md. A. Shoeb , Md. Zakir Hossain

**Abstract**— Internet generation is growing accustomed to having broadband access wherever they go and not just at home or in the office, which turns mobile broadband into a reality. This paper aims to look for a suitable mobile broadband solution in the South Asian region through comparative analysis in various perspectives. Both WiMAX and LTE are 4G technologies designed to move data rather than voice having IP networks based on OFDM technology. Proving competency in various significant aspects WiMAX and LTE already have made a strong position in telecommunication industry. Again, because of certain similarities in technology; they aren't like technological rivals as of GSM and CDMA. But still they are treated as opponents and viewed as a major threat in case of the flourishing of each other. Such view point is surely not conducive for getting the best out of them. In this paper various aspects and applications of WiMAX and LTE for deployment have been analyzed. South Asia being the residence of an enormous number of people presents an exciting opportunity for mobile operators, developers and internet service providers. So, every consideration that has been made here also correlates successfully with south Asia i.e. how mass people of this region may be benefited from it. As a result, it might be regarded as a good source in case of making major BWA deployment decisions in this region. Besides these, it also opens the path for further research and thinking in this issue.

——————————— ◆ ———————————

## 1 INTRODUCTION

Broadband wireless is a technological confluence in bringing the broadband experience to a wireless context, which offers users certain unique benefits and convenience. Broadband is going mass market while consumers are increasingly mobile. Mobile telephony leads the telecommunications market and data services represent an increasing share of total mobile ARPU.Thus, there is a crying need for deploying a cost-effective and scalable wireless broadband technology in this region to meet the broadband hunger of the classes as well as the masses [1].

South Asia, also known as Southern Asia, is the southern region of the Asian continent typically consisting of Bangladesh, Bhutan, India, the Maldives, Nepal, Pakistan and Sri Lanka. Some definitions may also include Afghanistan, Myanmar, Tibet, and the British Indian Ocean Territories. South Asia is home to well over one fifth of the world's population, making it both the most populous and most densely populated geographical region in the world [8].So, looking for an ideal mobile broadband solution for the mass population in this region is really a vital decision and thus requires analyzing from various point of view. Such is the approach of this paper.

WiMAX stands for Worldwide Interoperability for Microwave Access. It is a 4th generation cellular telecommunication technology currently based on IEEE 802.16e standard. Mobile WiMAX based on IEEE 802.16e-2005 [2] standard is an amendment of IEEE STD 802.16-2004 [3] for supporting mobility [28] [29]. IEEE 802.16-2004 is also frequently referred to as "fixed WiMAX" since it has no support for mobility [3].The IEEE 802.16m [4] standard is the core technology for the proposed Mobile WiMAX Release 2, which enables more efficient, faster, and more converged data communications.

Long Term Evolution (LTE) offers a superior user experience and simplified technology for next-generation mobile broadband. LTE is the next major step in mobile radio communications and is introduced in 3GPP (3rd Generation Partnership Project) Release 8. It is the last step toward the 4th generation (4G) of radio technologies designed to increase the capacity and speed of mobile telephone networks [5].The world's first publicly available LTE-service was opened by TeliaSonera in the two Scandinavian capitals Stockholm and Oslo on the 14th of December 2009 [6][7].

In this paper, an attempt has been made to facilitate a planned decision making stage for the mobile broadband solution specifically focusing in the South Asian region.

————————————————

- *Nafiz Imtiaz Bin Hamid is with the Department of Electrical and Electronic Engineering (EEE), Islamic University of Technology (IUT), Board Bazar, Gazipur-1704, Bangladesh.*
- *Md. R. H. Khandokar is with the School of Engineering and Computer Science. Independent University, Bangladesh.*
- *Taskin Jamal is with the Electrical and Electronic Engineering Department, University of Asia Pacific, Dhanmondi, Dhaka-1209, Bangladesh.*
- *Md. A. Shoeb is with the Electrical and Electronic Engineering Department, Stamford University, Siddeswari, Dhaka-1217, Bangladesh.*
- *Md. Zakir Hossain is with the Radio Access Network (RAN) Department, Qubee. Augure Wireless Broadband Bangladesh Limited.*

## 2 MOBILE BROADBAND

### 2.1 Why Mobile Broadband

- Mobile broadband provides hassle-free anytime, anywhere connectivity. There are terrible traffic conditions in various parts of South Asia. Long hours in the car means that everyone wants to find something to do in rush hour, and mobile Internet meets the demand.
- Millions of people continue to pay for line rental on a home phone that they rarely use, just so that they can get broadband at home. But with mobile broadband, there's no need to pay line rental leading towards a great save.
- Exciting daily package available when one is on the go or in roaming.
- Mobile broadband is mostly a plug and play technology. All the software needed to connect to the internet is automatically installed during the first time of connection giving the consumer an easier experience of internet usage.
- Unreliable electricity supply in the developing countries (which is also a common concern in South Asia) limits personal computer usage and consumer's ability to enjoy fixed line broadband.

Mobile broadband is thus expected to achieve significant penetration levels in developing/emerging economies in the relatively near future [30] [31].

### 2.2 Growth of Mobile Broadband

There is strong evidence supporting predictions of increased mobile broadband usage. Consumers understand and appreciate the benefits of mobile broadband. Broadband subscriptions are expected to reach 3.4 billion by 2014 and about 80 percent of these consumers will use mobile broadband shown in Fig 1 [9] [32].

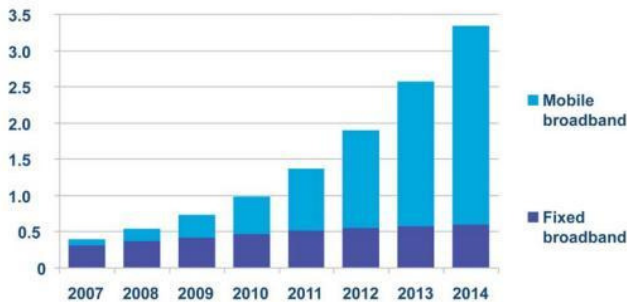

Fig. 1. Fixed and mobile broadband growth 2007-2014

Mobile broadband is becoming a reality, as the internet generation grows accustomed to having broadband access wherever they go and not just at home or in the office. Of the estimated 3.4 billion people who will have broadband by 2014, about 80 percent will be mobile broadband subscribers [9] [32].

## 3 CURRENT NETWORK DEPLOYMENT SCENARIO

Although the hype around WiMAX is quickly dissipating, still the standard has gained enough backing and volume to serve as an alternative for the provisioning of mobile broadband access. It has begun to carve out a tight niche tied to certain target opportunities, it has inspired a new wireless business model, and it has a flexible, flat, all-IP network architecture better suited than HSPA to providing Internet-based services. In contrast, however, the LTE standard has quickly gained substantial momentum. Since WiMAX 802.16e and LTE release 8 will provide similar real-world performance, ultimately the decisions of the largest WiMAX players may determine the fate of WiMAX [10].

The number of WiMAX deployments — currently more than 500 across 145 countries is greater than that of any conventional 3G technology and more than 50% greater than the number of HSPA network commitments. However, most WiMAX deployments to date have been small and expected to increase the coverage. Many of the larger WiMAX deployments are still underway, and large countries such as India, Indonesia and Vietnam are just beginning to issue WiMAX licenses [10].

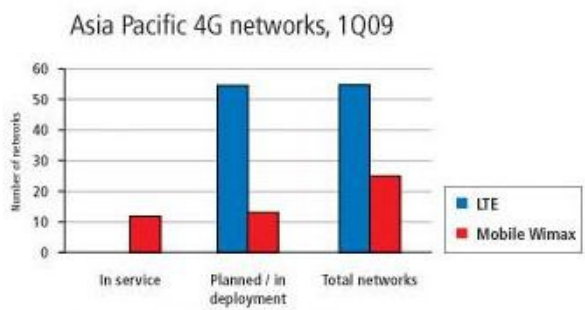

Fig. 2. Comparing LTE with WiMAX in Asia Pacific 4G Networks

TelecomAsia recently reported in Asia Pacific there were 55 LTE networks in these categories compared to 25 mobile WiMAX networks in 1Q09, although mobile WiMAX had the clear advantage of 12 networks in service compared to zero for LTE, according to Informa Telecoms & Media research.[11].The fact is shown in Fig 2.

## 4 DEPLOYMENT IN SOUTH ASIA-A CASE STUDY

South Asia is home to well over one fifth of the world's population. Being such a densely populated developing region in the world map; the answer of ideal mobile broadband solution for this zone is quite a challenging task. Because, there remains a huge difference in the life style led by the different classes of people in this region. So, when the question of choosing the perfect mobile broadband solution comes. It must have to be economically suitable to the general people. Let's now take a quick look on the current status of different south Asian countries in case of mobile broadband deployment [32]:

## 4.1 WiMAX Network

- **India:**

Though India does not have any B2C WiMAX network, a few companies are doing on experimental basis. BSNL,India successfully launched WiMAX in Goa in July,2009 along with Gujrat with a future target on Kerala and Punjab.BSNL has 20 MHz of BWA spectrum in 2.5 Ghz frequency.They already completed the initial deployment of 1000 BS and planning to launch more quickly [8][14].Except BSNL, other contributions came through Aircell through its Business Solution (ABS) plan .It is the first company in India to launch WiMAX and one among the five global operators to achieve this feat [8].They initiated their dominance through WiMax-ing Chennai. Again, Reliance launched servics on Pune, Bangalore on July 4, 2009 and VSNL Tata Indicom is also in the way of launching service [8].

- **Pakistan:**

Wi-Tribe Pakistan bringing together two shareholders Qatar Telecom (Qtel) and Saudi-based A.A started commercial services in June 2009 in 4 major cities, while expansion is underway for another 10 cities.Qubee has launched services for residential and business customers in Karachi. Super Broadband service is launching soon with testing network running in Karachi. Mobilink Infinity Pakistan's Second WiMAX network has operations in the South region currently [8]. Wateen Telecom headquartered in Lahore has successfully deployed one of the biggest nationwide WiMAX networks with 42 MHz of spectrum. It started commercial launch in December,2007; now deployed more than 842 four sector base stations across 22 cities with network covering 20% of Pakistan's 164 million inhabitants [8] [14].

- **Bangladesh:**

In Bangladesh, Agni- A Public Limited Company providing Internet Access in Bangladesh since 1995; launched Pre-WiMAX in 2006. Later on they started deploying Fixed WiMAX on 3.5 GHz Band before the arrival of mobile broadband. Then the era of mobile broadband started with Qubee from Augere with a mission "Broadband for all". They launched wireless broadband internet services for residential and business customers in Dhaka in October 2009 [8] [12]. Banglalion acquiring the 1st BWA license from Bangladesh Telecommunication Regulatory Commission (BTRC) on November 18, 2008 and committed to provide Broadband Internet connectivity and other services using WiMAX technology. By the 2nd quarter of 2010, Banglalion plans to bring the whole Dhaka city and its wider periphery under WiMAX coverage. By the end of 2nd quarter 2011, Banglalion plans to bring most areas of the country under its seamless coverage [13].

- **Srilanka:**

In Srilanka, Dialog Telekom began commercial operations in late 2006 and offers speeds of up to 4 Mbit/s. Sri Lanka Telecom has also launched test transmission in certain areas.Lanka Bell launched commercial operations in early 2008.Again,SUNTEL has started the Wi-Max BroadBand [8].

## 4.2 LTE Network

In December 2009, TeliaSonera became the first operator in the world to offer commercial 4G/LTE services, in the central parts of Stockholm and Oslo; network being supplied by Ericsson and Huawei [15] [8].

But reality for South Asia isn't the same. Potential network opening time for this region is way too far from the above mentioned one. Up to now in the South Asian region only India and Pakistan have some definite plan for LTE deployment. But it isn't before quarter 4 of 2012.The following chart shows the deployment plan for the south Asian countries for possible deployment of LTE [16].

TABLE 1

POSSIBLE DEPLOYMENT PLAN OF LTE IN SOUTH ASIAN REGION

| Country | Operator | Potential Opening Time |
|---|---|---|
| India | BSNL | Q4 2012 |
| Pakistan | PMCL | December 2014 |
|  | Telenor | March 2014 |

But luckily in some extent, it gives a scope for us to rethink the overall network deployment plan for this region [32].

## 5 TECHNICAL ASPECTS OF WIMAX AND LTE

Currently deployed network uses WiMAX standard IEEE 802.16e for ensuring mobility. Future ratification of the current standard is IEEE 802.16m with improved features and higher data rates where the network update procedure is also possible from the current one [4] [8] [18] [19]. LTE will take time to roll out, with deployments reaching mass adoption by 2012 while WiMAX is out right now.

The following table, Table 2 [17] [32] shows a comparison between the currently deployed IEEE 802.16e and 3GPP LTE-EUTRAN.

## 6 IEEE 802.16M-A FUTURE DEVELOPMENT IN WIMAX STANDARD

The IEEE 802.16m standard is the core technology for the proposed Mobile WiMAX Release 2 [8]. Among many enhancements, IEEE 802.16m systems can provide four times faster data speed than the current Mobile WiMAX Release 1 based on IEEE 802.16e technology. It can push data transfer speeds up to 1 Gbit/s while maintaining backwards compatibility with existing WiMAX radios [4][8]. IEEE 802.16m shall meet the cellular layer requirements of IMT-Advanced next generation mobile networks [20].

LTE is also trying to push mobile broadband to a theoretical 100 Mbits/second for downlink and 50 Mbits/s for uplink. It is well beyond what HSPA will be able to offer in a year or so.Cellular's LTE deployment is expected between late 2009 and 2011, the same time frame as the newly proposed 802.16m Mobile WiMAX standard [4].

TABLE 2

IEEE 802.16E & 3GPP-LTE (E-UTRAN) - A COMPARISON

| Aspect | Mobile WiMAX (IEEE802.16e-2005) | 3GPP-LTE (E-UTRAN) |
|---|---|---|
| Core network | WiMAX Forum™ All-IP network | UTRAN moving towards All-IP Evolved UTRA CN with IMS |
| Access technology: Downlink (DL) Uplink (UL) | OFDMA OFDMA | OFDMA SC-FDMA |
| Frequency band | 2.3-2.4GHz, 2.496-2.69GHz, 3.3-3.8GHz | Existing and new frequency bands (~2GHz) |
| Bit-rate/Site: DL UL | 75Mbps (MIMO 2TX 2RX) 25Mbps | 100Mbps (MIMO 2TX 2RX) 50Mbps |
| Channel bandwidth | 5, 8.75, 10MHz | 1.25-20MHz |
| Cell radius | 2-7Km | 5Km |
| Cell capacity | 100-200 users | >200 users @ 5MHz >400 users for larger BW |
| Spectral efficiency | 3.75[bits/sec/Hz] | 5[bits/sec/Hz] |
| Mobility: Speed Handovers | Up to 120Km/H Optimized hard handovers supported | Up to 250Km/H Inter-cell soft handovers supported |
| Legacy | IEEE802.16a through 16d | GSM/GPRS/EGPRS/UMTS/HSPA |
| MIMO: DL UL No. of code words | 2Tx X 2Rx 1Tx X NRx (Collaborative SM) 1 | 2Tx X 2Rx 2Tx X 2Rx 2 |
| Standardization coverage | IEEE 802.16e-2005 PHY and MAC CN standardization in WiMAX forum™ | RAN (PHY+MAC) + CN |
| Roaming framework | New (work in process in WiMAX Forum™) | Auto through existing GSM/UMTS |
| Schedule forecast: Standard completed Initial Deployment Mass market | 2005 2007 through 2008 2009 | 2007 2010 2012 |

TABLE 3

AIR INTERFACE REQUIREMENT FOR IEEE 802.16M SPECIFICATION

| Characteristic | | Requirement |
|---|---|---|
| Peak Data Rate | Downlink | 1Gbps @ Nomadic 100Mbps @ High mobility |
| | Uplink | TBD |
| Expected spectral efficiency | Micro cell (DL/UL) | TBD |
| | Macro cell (DL/UL) | TBD |
| Bandwidth | | Scalable bandwidth including 5, 7, 8.75, 10 MHz |
| Center frequency | | Frequency is expected to be decided in WRC07 |

Table 3 shown above [20] [32] elaborates the proposed air interface requirement for this IEEE 802.16m standard. The 802.16m profile is currently under evaluation and is expected to be ratified along with WiMAX Release 2 later this year. The first 802.16m dongles is expected to be visible in late 2011 while more wide-spread commercial deployments to be starting in 2012 [21].

## 7 LTE-ADVANCED

### 7.1 Overview

LTE-Advanced extends the technological principles behind LTE into a further step change in data rates. Incorporating higher order MIMO (4x4 and beyond) and allowing multiple carriers to be bonded together into a single stream, target peak data rates of 1Gbps have been set. It also intends to use a number of further innovations including the ability to use non-contiguous frequency ranges, with the intent that this will alleviate frequency range issues in an increasingly crowded spectrum, self back-hauling base station and full incorporation of Femto cells using Self-Organizing Network techniques [25].

Some of the main aims for LTE-Advanced are shown below [26] [32] in Table 4

TABLE 4

LTE-ADVANCED: SOME MAIN AIMS

| Peak data rates | DL:1Gbps |
| | UL:500Mbps |
| Spectrum efficiency | 3 times greater than LTE |
| Peak spectrum efficiency | DL:30 bps/Hz |
| | UL:15 bps/Hz |
| Spectrum use | The ability to support scalable bandwidth use and spectrum aggregation where non-contiguous spectrum needs to be used |
| Latency | From Idle to Connected in less than 50 ms and then shorter than 5 ms one way for individual packet transmission |
| Cell edge user throughput | Twice that of LTE |
| Average user throughput | 3 times that of LTE |
| Mobility | Same as that in LTE |
| Compatibility | LTE Advanced shall be capable of interworking with LTE and 3GPP legacy systems |

### 7.2 LTE-Advanced Compared to other 3G Services

Questions may arise that why LTE should come into play where there are other 3G services working in various parts. The answer is the effectivity of the technology with its associated advantage. The development of LTE Advanced/IMT Advanced can be seen as an evolution from the 3G services that were developed using UMTS / W-CDMA technology. A comparison [26] is shown below in Table 5.

TABLE 5

LTE ADVANCED, LTE AND OTHER 3G TECHNOLOGIES

| | WCDMA (UMTS) | HSPA HSDPA / HSUPA | HSPA+ | LTE | LTE ADVANCED (IMT ADVANCED) |
|---|---|---|---|---|---|
| Max downlink speed bps | 384 k | 14 M | 28 M | 100M | 1G |
| Max uplink speed bps | 128 k | 5.7 M | 11 M | 50 M | 500 M |
| Latency round trip time approx | 150 ms | 100 ms | 50ms (max) | ~10 ms | less than 5 ms |
| 3GPP releases | Rel 99/4 | Rel 5 / 6 | Rel 7 | Rel 8 | Rel 10 |
| Approx years of initial roll out | 2003 / 4 | 2005 / 6 HSDPA 2007 / 8 HSUPA | 2008 / 9 | 2009 / 10 | |
| Access methodology | CDMA | CDMA | CDMA | OFDMA / SC-FDMA | OFDMA / SC-FDMA |

## 8 FEW VITAL FACTORS

### 8.1 Backward Compatibility

According to the announcement of WiMAX Forum, Wi-MAX Release 2, which is based on the standard, would be finalized in parallel with 802.16m.It is just to ensure that the next generation of WiMAX networks and devices will remain backward compatible with WiMAX networks based on 802.16e [21].

Samsung's Mobile WiMAX solution provides strong backward compatibility with Mobile WiMAX Release 1 solutions. It allows current Mobile WiMAX operators to migrate their Release 1 solutions to Release 2 with little expense by upgrading channel cards and their base station software. Also, the subscribers who use currently available Mobile WiMAX devices can communicate with new Mobile WiMAX Release 2 systems without difficulty [22].

LTE has the advantage of being backwards compatible with existing GSM and HSPA networks, enabling mobile operators deploying LTE to continue to provide a seamless service across LTE and existing deployed networks [25].

### 8.2 Spectrum Issue

Most 3G networks operate using up to 5MHz channels, WiMAX 802.16e networks operate using up to 10MHz, and 802.16m and LTE networks will operate using up to 20MHz channels. To achieve the significantly higher performance as reported by TeliaSonera, LTE operators need to use the wider 20MHz channels, but that spectrum is not always readily available [21].Lots of the spectrum allocation are in 10MHz chunks and the places with contiguous 20MHz channels are few and far between according to Intel.

So, this spectrum issue might add to complexity for the overall satisfactory performance of LTE.LTE-Advanced bearing 3GPP Release 10 may bring some relief in this regard.

## 9 CURRENT GROWTH AND POSSIBLE FUTURE OUTCOME

### 9.1 WiMAX

New Intel Core Series notebook with integrated Wi-Fi/WiMAX module can easily facilitate these issues [24]. Again, cellular phone with GSM/WiMAX dual mode is about to come; it also can smooth the scenario towards more improvement. HTC already announced the HTC MAX 4G device that is the world's first GSM/WiMAX headset [27][32].

### 9.2 LTE

Samsung isn't just prepping WiMAX release 2 (802.16m) for 4G deployment but also develop its LTE USB modem. Samsung GT-B3710, the First LTE USB Modem Achieved Interoperability from Ericsson LTE Network in Stockholm. LG is also working dominantly in this respect by recently unveiling its LTE USB modem at CommunicAsia 2009 and showing up another model at CEATEC JAPAN 2009 [6][32].

## 10 LITERATURE REVIEW

South Asia though having an enormous population is still considered as a neglected, developing region. So, works based on South Asia is relatively less. In case of finding suitable mobile broadband solution we couldn't find any related work with the best of our effort. But it is of great importance as it may change the fate of mass population in this region.

[2] & [3] are the IEEE standards for fixed and mobile WiMAX. [4],[18],[20],[21] & [22] gives the technical overview of IEEE standard 802.16m which is supposed to work in parallel with mobile WiMAX release 2.A suitable mobility handover scenario has been depicted in [28] & [29].[10],[11],[17],[19],[30],[31] & [34] analyze the general notion about the WiMAX and LTE and obstacles behind their proper flourishment. They also try to break the wrong and unfriendly interrelations between these two technologies. A pathway for eradicating conflicts and thus implementing better mobile broadband has been elaborated in [32].

Other references like [33] reflect various up to date scenarios in case of network deployment or device arrivals. Those have been picked to strengthen analysis in this paper.

## 11 SOME ATTRACTIVE APPLICATIONS OF WIMAX WITH REGIONAL CONSIDERATION

### 11.1 Disaster Management

WiMAX access was used to assist with communications in Aceh, Indonesia, after the tsunami in December 2004 when all communication infrastructure in the area, other than amateur radio, was destroyed, making the survivors unable to communicate with people outside the disaster area and vice versa. Then WiMAX provided broadband access that helped regenerate communication to and from

Aceh.

In addition, WiMAX was donated by Intel Corporation to assist the FCC and FEMA in their communications efforts in the areas affected by Hurricane Katrina. In practice, volunteers used mainly self-healing mesh, VoIP, and a satellite uplink combined with Wi-Fi on the local link [8] [32].

♦ **Regional Perspective**

In the South Asian region, we often find the natural calamities like flood, cyclone, earthquake and people being affected. For the disaster management plan mobile broadband solution like WiMAX can contribute a greater role if it can be deployed successfully all over the country [32].

## 11.2 Providing Public Safety

Through WiMAX, public safety agencies can be connected with each other. During any mishap, such as accident, fire etc., the control office can send its command to the police station, hospital, or fire brigade office. The corresponding agencies immediately can connect to the accidental location by using WiMAX-enabled vehicles. The video images and data from the site of accidental location can be sent to corresponding agencies. These data can be examined by the experts of the emergency staff and accordingly prescription can be communicated. A video camera in the ambulance can send the latest images of the patient before the ambulance reaches the hospital so that the doctors can get ready for further action quickly. Through WiMAX, a fireman can download the data about the best possible route to a fire scene [33]. Such a symbolic scenario for ensuring public safety is shown [33] below in Figure 3.

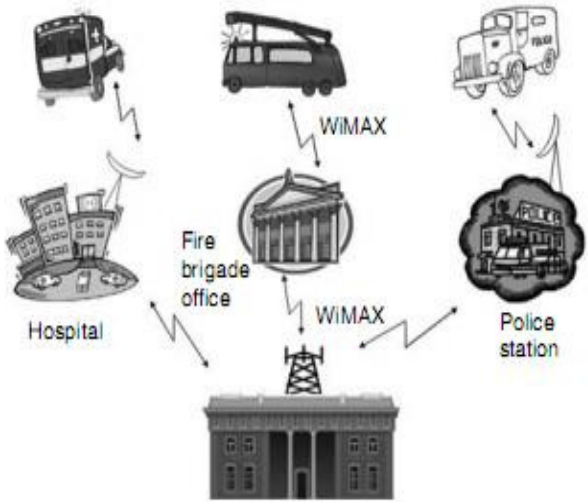

Fig. 3. Ensuring Public Safety through WiMAX Connectivity

♦ **Regional Basis**

In dry seasons, lots of accidents take place because of electrical short circuits. The outcome is the dangerous fire destroying the surroundings. Things are even more serious in case of the garments factory with inadequate fire extinguishing measures along with old unplanned shopping malls. If a planned mobile broadband network can be built covering this problem some zones, then things may be brought under light through the scenario mentioned under public safety.

## 11.3 Campus and Institutional connectivity

Campus system requires high data capacity, a large coverage along with high security. WiMAX can connect various blocks within the campus which can be very difficult through cables because the lead time to deploy a wired solution is much longer than the same deployment using a WiMAX solution. Such a network is shown in Figure 4.

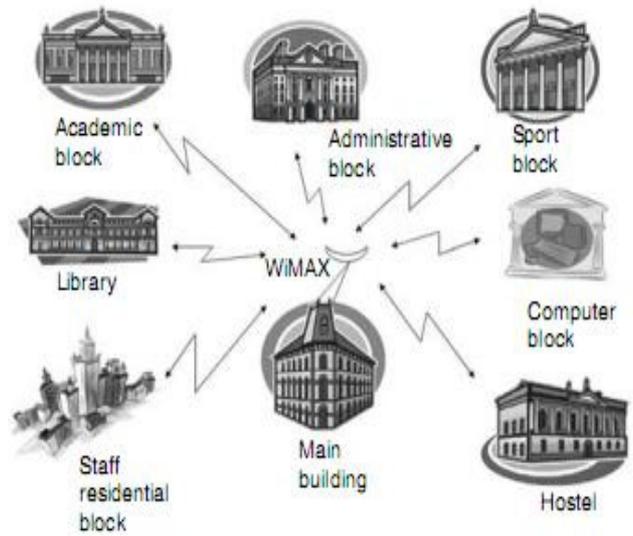

Fig. 4. Campus connectivity using WiMAX

WiMAX can thus connect boards, colleges, schools, and the main head offices.Through this, telephone voice, data, email, Internet, question papers, intranet, video lectures, presentations, and students' results can be communicated at a very high rate. By video conferencing the students can interact with the teachers of another institution. A camera at campus 1 of a university may deliver real-time classroom instruction to campus 2, allowing the colleges to simultaneously deliver instruction from a recognized subject matter expert to a large number of students. College and schools in rural areas can be connected through WiMAX with other institutions having better facilities through WiMAX so that remotely located students can also be benefited. One typical scenario has been shown in Figure 5.

♦ **Regional Case-Study**

This can be a substantial solution for the developing countries like those in the South Asian region.Again, standard of education can also be raised to equity through this.

## 11.4 Traffic Control/ Vehicle Tracking

The availability of an all IP, mobile, high-speed network has also generated some innovative applications. To better help students track the location of shuttle buses, Ball State university- a cutting-edge wireless research univer-sity in Muncie, Indiana; installed notebook computers equipped with WiMAX and GPS USB dongles in the front panels of the busses. The information is then reported back to a fleet management server, providing real-time location based information that can be accessed by the students [23] [32].

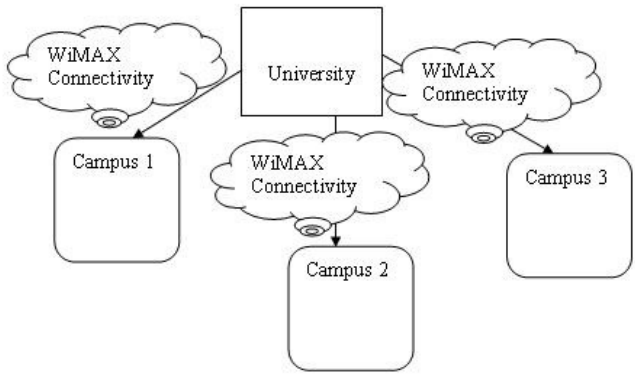

Fig. 5. Distance learning through online education

- **Regional Scenario-South Asia**

South Asia is enormously being affected by poor traffic control system. In this region, to ensure a happy organized life of people maintaining a planned traffic control system is deadly necessary. But we are finding that lots of plans being initiated are going in vain like the case of Bangladesh. In this regard, we can think of:

- Building a real-time location based system employing WiMAX which is the available mobile broadband solution option right now. Later on we can think of utilizing LTE or LTE-Advanced in this regard.
- Planned time information based bus/train service with appropriate arrival time for the passengers may be provided with this in this region.
- Again, for security issue like car, taxi theft vehicle tracking system is deadly necessary. Mobile broadband can facilitate this.

## 12 KEY FINDINGS & FUTURE CHALLENGE

WiMAX and LTE aren't the technological rivals like GSM and CDMA.Rather with the future upgardation of current standard of WiMAX IEEE 802.16e to IEEE 802.16m; they will tend to provide quite similar features. But it can't be denied that WiMAX has got a certain advantage of earlier deployment in this wireless field. But still LTE will start in South Asia most possibly at the end of 2012 [16].Again, it has also an excellent update i.e. 3GPP Release 10, LTE-Advanced [26].So, if they are deployed as they chronologically arrive in the telecom field with no rivalry among the network vendors and operators; we can surely hope for getting the perfect broadband solution for subscribers in this region.Again, through this way of approaching, we can ensure the arrival of two quite similar technologies based on performance around 2012.It will obviously increase the choice for consumers in quality mobile broad-band solution. So, mobile broadband experience will be at its possible level best [32].

- **Transition from TDD to FDD and vice versa**

Both LTE and WiMAX are based on OFDM technology. As a result, it shouldn't be an unwise decision to think of chipsets supporting both WiMAX and 3GPP LTE. That is the support of both TDD & FDD through the same baseband chip. Such an idea has already arisen in the mind of Wavesat in their chip Odyssey 9010 enabled for the switching from TDD to FDD. Again, the Beceem 4G-LTE/WiMAX chip that supports up to 150-Mbit/s downlink speeds has also the ability to connect to any 4G LTE or WiMAX network with seamless roaming, and switching between TDD and FDD configuration as needed [34].It frees operators from concerns of how to best use their valuable spectrum.

The ultimate outcome might be:
- Support for IEEE 802.16m along with LTE protocol stack including MAC, RLC, PDCP, RRC, and NAS layers; turning the package like both LTE and Wi-MAX in a system-on-a-chip (SoC).
- Enabling real-time band or channel reconfiguration through a unique multi-mode "autosense" feature which can automatically detect the network type.
- End or at least mitigation of the so-called battle between Long-Term Evolution (LTE) and WiMAX for 4G cellular service dominance.

## 10 CONCLUSION

As backward compatibility can be ensured in IEEE 802.16m through easier upgrading from IEEE 802.16e; WiMAX should never lag awkwardly behind LTE or LTE-Advanced. Again, earlier deployment should give it the advantage over LTE. But it shouldn't create the rivalry in between them. Because, the overall technologies used by them are mostly same .So, no how they should be treated as rivals. Now South Asia has WiMAX deployment in some countries which are likely to spread. But LTE deployment should take time and possibly LTE won't enter in South Asia before Q4 2012.But mobile broadband being an enormously popular technology with varied features, use of existing technology i.e. mobile WiMAX right now should be encouraged. Otherwise effective penetration won't take place through the step by step upgrade WiMAX Release 2 (IEEE 802.16m).In the same way then, LTE won't be able to get the dazzling upgrade LTE-Advanced. Rather we should ensure the best possible utilization of these technologies in various aspects. In this paper these issues have been elaborated suitably with respect to South Asia perspective and the current scenario. Significant utilizations of these technologies have been put in light with the regional consideration. And a possible prospective solution path towards finding the best possible mobile broadband experience for subscribers has been depicted.

## ACKNOWLEDGMENT

The authors wish to thank Mr. Rokibul Islam Rony, Wi-MAX Engineer, Huawei Technologies Bangladesh Ltd.

## REFERENCES

[1] Fundamentals of WiMAX: Understanding Broadband Wireless Networking by Jeffrey G. Andrews, Arunabha Ghosh, Rias Muhamed
[2] IEEE STD 802.16e-2005. "IEEE Standard for Local Metropolitan Area Networks- Part 16: Air Interface for Fixed & Mobile Broadband Wireless Access Systems", 28 February 2006.
[3] IEEE STD 802.16-2004, "Air Interface for Fixed Broadband Wireless Access Systems", October 2004.
[4] WiMAX 802.16m: 1Gbps
http://www.dailywireless.org/2007/02/20/wimax-80216m-100-mbps/
[5] LTE – The UMTS Long Term Evolution From Theory to Practice by Stefania Sesia, Issam Toufik, Matthew Baker.
[6] Long Term Evolutionist LTE News, LTE Deployment, LTE Devices, LTE Operators.
http://www.longtermevolutionist.com/
[7] LTE for UMTS – OFDMA and SC-FDMA Based Radio AccessEdited by Harri Holma and Antti Toskala.
[8] Wikipedia-the free Encyclopedia
http://en.wikipedia.org/wiki/
[9] White paper:LTE-an introduction by Ericsson
[10] WiMAX and LTE: The Case for 4G Coexistence
http://www.wimax.com/commentary/news/wimax_industry_news/2010/january-2010/wimax-and-lte-the-case-for-4g-coexistence-0106
[11] LTE vs WiMAX in Asia Pacific
http://www.longtermevolutionist.com/2009/09/28/lte-vs-wimax-in-asia-pacific/
[12] QUBEE
http://www.qubee.com.bd/
[13] Banglalion Communications Ltd
http://www.banglalionwimax.com/
[14] WiMAX: the Quintessential Answer to Broadband to India by protivity
[15] The World's First 4G/LTE Speedtest on TeliaSonera in Stockholm
http://www.longtermevolutionist.com/2010/01/23/the-worlds-first-4glte-speedtest-on- teliasonera-in-stockholm/
[16] LTE Commitments September 2009
www.3gamericas.org/.../LTE%20Commitments%20September%202009.pdf
[17] A Comparison of Two Fourth generation Technologies: WiMAX and 3GPP-LTE by Jacob Scheim
[18] WiMAX Act 2: 802.16m Provides Evolution Path to 4G
http://www.wimax.com/commentary/blog/blog-2010/february-2010/wimax-act-2-80216m-provides-evolution-path-to-4g-0203
[19] LTE vs WiMAX: A Little 4G Sibling Rivalry
http://gigaom.com/2008/03/05/a-little-4g-sibling-rivalry/
[20] IEEE 802.16m Requirements
IEEE 802.16 Broadband Wireless Access Working Group <http://ieee802.org/16>
[21] IEEE 802.16m Update by Raj Jain, Washington University in Saint Louis
[22] Samsung's Simplest Migration to Mobile WiMAX Release 2
http://www.wimaxian.com/2009/10/12/samsungs-simplest-migration-to-mobile-wimax-release-2/
[23] Wireless Broadband Perspectives - WiMAX.com Weekly Series Sponsored by Cisco
http://www.wimax.com/commentary/blog/blog-2010/january-2010/opportunity-knocks-us-universities-look-to-capitalize-on-mobile-wimax-0119
[24] CES 2010: WiMAX Wrap-up
http://www.wimax.com/commentary/blog/blog-2010/january-2010/ces-2010-wimax-wrap-up-0112
[25] LTE
http://www.gsmworld.com/technology/lte.htm
[26] LTE Advanced for IMT 4G
http://www.radio-electronics.com/info/cellulartelecomms/4g/3gpp-imt-lte-advanced-tutorial.php
[27] HTC announces world's first GSM/WiMAX mobile phone
http://blogs.zdnet.com/cell-phones/?p=249
[28] Mohammad Kawser, Nafiz Imtiaz Bin Hamid, Mohammad Abu Naser and Adnan Mahmud, "An Optimized Method for Scheduling Process of Scanning for Handover in Mobile WiMAX"- International Conference on Computer and Network Technology (ICCNT 2009),Chennai, India.
[29] Nafiz Imtiaz Bin Hamid and Adnan Mahmud, "Optimizing the Handover Procedure in IEEE 802.16e Mobile WiMAX Network"- 3rd National Conference on Communication and Information Security (NCCIS 2009), Dhaka, Bangladesh. pp.23-28
[30] Mobile broadband vs. fixed line broadband
http://www.broadbandchoices.co.uk/mobile-broadband-vs-fixed-line-broadband.html
[31] Mobile broadband for the masses: Regulatory levers to make it happen by McKinsey & Company
[32] Nafiz Imtiaz Bin Hamid, Md. Zakir Hossain, Md. R. H. Khandokar, Taskin Jamal and Md.A. Shoeb -"Mobile Broadband Possibilities considering the Arrival of IEEE 802.16m & LTE with an Emphasis on South Asia" -International Journal of Computer Science and Information Security (IJCSIS), Vol. 7 No.3, March 2010. pp. 267-275. ISSN: 1947-5500.
[33] WiMAX Applications by Syed Ahson and Mohammad Ilyas
[34] LTE Or WiMAX? How About Both?
http://mobiledevdesign.com/standards_regulations/lte-wimax-about-both-040810/

**Nafiz Imtiaz Bin Hamid** received his Bachelors degree in Electrical and Electronic Engineering from Islamic University of Technology(IUT) in 2008 and now pursuing his M.Sc. Engineering degree.He has been working as a lecturer in the Electrical and Electronic Engineering Department of IUT since 2009.His primary research interest includes Broadband Wireless Access (BWA) i.e. PHY/MAC layer protocol along with mobility related issue analysis of 4G cellular technologies like WiMAX,LTE etc.Nafiz is a graduate student member of IEEE and ACM.He is also a member of IACSIT.Nafiz is included in the Technical Program Committee of IEEE ISIEA 2010 & CSSR 2010 to be held in Malaysia. He is also the reviewer of several peer-reviewed International Journals.

**Md. R. H. Khandokar** received his Bachelors degree in Electrical and Electronic Engineering from Islamic University of Technology (IUT) in 2008.Currently he is serving as a lecturer in the School of Engineering and Computer Science (SECS) of Independent University, Bangladesh.

**Taskin Jamal** received his Bachelors degree from Electrical and Electronic Engineering department of Islamic University of Technology (IUT) in 2008.Currently he is working as a lecturer in the Electrical and Electronic Engineering department of The University of Asia Pacific (UAP), Bangladesh.

**Md. A. Shoeb** received his Bachelors degree from Electrical and Electronic Engineering department of Islamic University of Technology (IUT) in 2008.He is now pursuing his M.Sc. Engineering degree from Bangladesh University of Engineering and Technology (BUET).Again, he is also serving as the lecturer in Electrical and Electronic Engineering Department of Stamford University, Bangladesh.

**Md. Zakir Hossain** received his Bachelors degree in Electrical and Electronic Engineering from Islamic University of Technology (IUT) in 2008. Currently he is working in the Radio Access Network (RAN) department of QUBEE, Augere Wireless Broadband Bangladesh Limited.